\documentclass[10pt]{article}
\usepackage{fullpage}
\usepackage{amsmath}
\usepackage{amssymb}
\usepackage[dvips]{epsfig}
\usepackage{color}

\def\##1{\underline #1}
\def\=#1{\underline{\underline #1}}

\def\eps{\epsilon}
\def\epso{\epsilon_0}

\def\.{\mbox{ \tiny{$^\bullet$} }}

\def\ux{\#{u}_x}
\def\uy{\#{u}_y}
\def\uz{\#{u}_z}

\def\un{\#{u}_n}
\def\ut{\#{u}_\tau}
\def\ub{\#{u}_b}

\def\le{\left(}
\def\ri{\right)}
\def\les{\left[}
\def\ris{\right]}
\def\lec{\left\{}
\def\ric{\right\}}

\def\c#1{\cite{#1}}
\def\l#1{\label{#1}}
\def\r#1{(\ref{#1})}

\begin{document}
\begin{center}

{\bf {\LARGE Determination of constitutive and morphological
parameters of columnar thin films by inverse homogenization}}

\vspace{10mm} \large

 Tom G. Mackay\footnote{E--mail: T.Mackay@ed.ac.uk.}\\
{\em School of Mathematics and
   Maxwell Institute for Mathematical Sciences\\
University of Edinburgh, Edinburgh EH9 3JZ, UK}\\
and\\
 {\em NanoMM~---~Nanoengineered Metamaterials Group\\ Department of Engineering Science and Mechanics\\
Pennsylvania State University, University Park, PA 16802--6812,
USA}\\
 \vspace{3mm}
 Akhlesh  Lakhtakia\footnote{E--mail: akhlesh@psu.edu}\\
 {\em NanoMM~---~Nanoengineered Metamaterials Group\\ Department of Engineering Science and Mechanics\\
Pennsylvania State University, University Park, PA 16802--6812, USA}

\end{center}

\vspace{4mm}

\normalsize
\begin{abstract}
A dielectric columnar thin film (CTF), characterized macroscopically
by a relative permittivity dyadic, was investigated theoretically
with the assumption that, on the nanoscale, it is an assembly of
parallel, identical, elongated ellipsoidal inclusions made of an
isotropic dielectric material that has a different refractive index
from the bulk material that was evaporated to fabricate the CTF. The
inverse Bruggeman homogenization formalism was developed in order to
estimate the refractive index of the deposited material, one of the
two shape factors of the ellipsoidal inclusions, and the volume
fraction occupied by the deposited material, from a knowledge of
relative permittivity dyadic of the CTF. A modified Newton--Raphson
technique was implemented to solve the inverse Bruggeman equations.
Numerical studies revealed how the three nanoscale parameters of
CTFs vary as functions of the vapour incidence angle.

\end{abstract}


\section{Introduction}

Columnar thin films (CTFs) are familiar structures within the optics
literature, having been fabricated by physical vapour deposition
methods for well over a century \c{HW}. Their morphology is
reminiscent of certain  crystals, while their macroscopic optical properties are
analogous to those of certain orthorhombic crystals.
Furthermore, they are the precursors of the more complex
sculptured thin films (STFs)  \c{STF_Book}.

The prospect of  controlling the porosity and the columnar
morphology of these thin films   at the fabrication stage, in order
 to engineer their macroscopic optical responses, renders them attractive
platforms for optically sensing chemical and biological species
\c{LMBR,L01,LMSWH,Steele,PH07,L07,{ML_OE},ML2008,PL2009}. However,
for intelligent design and deployment of such sensors, it is
important to fully characterize the relationship between macroscopic
constitutive properties on the one hand and the nanoscale morphology
and composition on the other.

There are significant impediments towards arriving at definitive
relationships. One is the variability that exists due to differences
in deposition conditions \cite{MTR1976,BMYVM}. For instance, the
bulk material that is evaporated may be quite different from the
material that is actually deposited as a thin film. Therefore, while
the dielectric properties of the bulk material is easily known prior
to evaporation, the dielectric properties of the deposited material
may well be different, depending on,  whether the deposition
occurred in an oxidizing or reducing atmosphere, whether trace
amounts of water vapor were present, and the temperature. As an
example, when Ti$_2$O$_3$ is evaporated, the deposited material has
been shown by one research group to be either TiO$_{1.8}$ or
TiO$_{1.5}$, depending on whether the temperature is 25~$^\circ$C or
250~$^\circ$C \cite{WRL03}. Evaporation of different suboxides of
titanium leads to the deposition of different TiO$_\alpha$ films, in
general, where the real number $\alpha$ varies with the nominal
deposition conditions and even the deposition apparatus.  Likewise,
when SiO$_2$ is evaporated, the deposited material is some
ill-defined but consistent mixture of Si and SiO$_2$ and is thus
often classified as SiO$_\alpha$, $\alpha\in(1,2)$ \cite[p.
164]{Macleod}. Furthermore, the delineation of nanoscale morphology
is not an  unambiguous task, as even a cursory glance at
scanning-electron-microscope images of CTFs will confirm \cite{HW}.
Direct determination of
porosity or void volume-fraction through a gas-adsorption technique
\cite{BET1938,BHBP,RHLP}, although accurate, is very time-consuming. Therefore, porosity is usually
measured
indirectly through measurement of mass density, which has its own
sources of inaccuracy  \cite{MTR1976}.

Various researchers  \c{HW,Smi89,Wang2,LW_Optik,Ward} have put forth nanoscale-to-macroscopic
models for the relative permittivity dyadics of CTFs.  Generally speaking, in  these models
the CTF is  viewed as an assembly of parallel, identical, nanoscale
inclusions of a certain shape dispersed in a certain homogeneous material. At optical and lower
frequencies, these inclusions are electrically small and can therefore be
homogenized into a macroscopically homogeneous material \cite{Mackay_JNP2008}. Apart
from the shape of the inclusions, one must choose the porosity and the bulk dielectric properties
of the deposited material and the material in the void regions (usually taken to be air) of the CTF.
Such models require careful calibration against experiments \cite{SLH}.

Inversion of the \emph{forward} homogenization
procedure can provide nanoscale information about a CTF, which can be useful,
for example, to predict what would happen if the CTF were to be infiltrated by some
other material \cite{Shalabney}. This thought motivated the work reported in this paper.

Provided the components of the relative permittivity dyadic of a CTF
are measured by suitable optical experiments \cite{HWH_AO,BZC},
 an \emph{inverse} homogenization
procedure could yield the refractive index of the deposited material, the porosity of the CTF,
and the shape of the inclusions, if certain reasonable assumptions are made. A demonstration
based on the   Bruggeman formalism  \cite{WLM}
is presented in
the following sections.
 In the notation adopted here, vectors are underlined
whereas dyadics are double underlined. The unit Cartesian vectors
are written as $\ux$, $\uy$, and $\uz$; the unit dyadic $\=I =
\ux \, \ux + \uy \, \uy + \uz \, \uz$; the permittivity of free space
is denoted by   $\epso$; the angular frequency is denoted by $\omega$;
and $i=\sqrt{-1}$.

\section{Homogenization model}

Let us consider a CTF  grown on a planar substrate through the
deposition of an evaporated bulk material. The planar substrate is
taken to lie parallel to the $z=0$ plane, and the deposited material
is assumed to be an isotropic dielectric material with refractive
index $n_s$.
 At length scales far greater than the nanoscale, the CTF is
effectively a continuum which may be characterized by the
frequency-domain constitutive relation  \c{HW,STF_Book}
\begin{equation}
 \#D  =  \epso\,\=\eps_{\,CTF}\. \#E\,, \label{epsbasic}
\end{equation}
where
\begin{equation}
\=\eps_{\,CTF}= \=S_{y} (\chi) \. \le \eps_{a} \,\uz\uz
+\eps_{b}\,\ux\ux \, +\,\eps_{c}\,\uy\uy \ri\.\=S_{y}^{-1}(\chi)\,
\label{ortho}
\end{equation}
is the relative permittivity dyadic of the CTF. The middle dyadic on the right side
of Eq.~(\ref{ortho}) indicates the macroscopic orthorhombic symmetry of the CTF \cite{HW}.
The orientation of the columns with respect to any $xy$ plane  is indicated via the inclination
dyadic
\begin{equation}
\=S_y(\chi) = \uy\uy + (\ux\ux + \uz\uz) \, \cos\chi +
(\uz\ux-\ux\uz)\, \sin\chi\,,
\end{equation}
where the column inclination angle is $\chi \in(0,\pi/2]$.

Each column of the CTF may be regarded as a set of elongated
ellipsoidal inclusions strung together end-to-end. All inclusions have the same orientation and shape.
The latter is specified through the shape dyadic
\begin{equation}
\=U_s = \un \, \un + \gamma_\tau \, \ut \, \ut + \gamma_b \, \ub \,
\ub,
\end{equation}
wherein the normal, tangential, and binormal basis vectors are specified
in terms of the column inclination angle per
\begin{equation}
\left. \begin{array}{l}
 \un = - \ux \, \sin \chi + \uz \, \cos \chi \vspace{4pt} \\
 \ut =  \ux \, \cos \chi + \uz \, \sin \chi \vspace{4pt} \\
\ub = - \uy
\end{array}
\right\}.
\end{equation}
Since the columnar morphology is highly aciculate, we have that the
shape parameters $\gamma_{b} \gtrsim 1$ and $\gamma_\tau \gg 1$.
As increasing $\gamma_\tau$ beyond
10 does not have significant effects for slender inclusions,
we fixed $\gamma_\tau = 15$ for definiteness.

As the CTF is   porous, we introduce  $f \in \le 0,1 \ri$
as the volume fraction occupied by the ellipsoidal inclusions
representing the columns of the CTF. The void region
is filled with air (or vacuum). Thus, the porosity of the CTF equals $1-f$.

\section{Forward and inverse homogenization}

The nanoscale parameters $\lec n_s,f,\gamma_b\ric$ may be related to
the eigenvalues $\lec \eps_a,\eps_b,\eps_c\ric$ of $\=\eps_{\,CTF}$
via one of several homogenization formalisms, including the Maxwell
Garnett formalism \cite{Ward}, the Bragg--Pippard formalism
\cite{HW}, and the Bruggeman formalism \cite{STF_Book}. We implement
here the last-named formalism which has been widely used in optics
\c{L96}, because it treats the region occupied by the deposited
material and the void region symmetrically, unlike the other two
formalisms.

Let us introduce  the dyadic
\begin{equation}
\label{Br} \=b = f \,\=a_{s}+(1-f) \,\, \=a_{f} \, ,
\end{equation}
which is the volume-fraction-weighted  sum of the two polarizability density dyadics
\c{STF_Book,Michel_chap}
\begin{eqnarray}
\nonumber \=a_{s} &=& \epso \les n_{s}^2\, \=I - \le\eps_{a}
\,\uz\uz +\eps_{b}\,\ux\ux
\,  +\,\eps_{c}\,\uy\uy\ri\ris \\
&&\. \lec \=I + i \omega \epso \,\=D_{s} \. \les n_{s}^2\, \=I -
\le\eps_{a} \,\uz\uz +\eps_{b}\,\ux\ux \, +\,\eps_{c}\,\uy\uy\ri\ris
\ric^{-1}\,
\end{eqnarray}
and
\begin{eqnarray}
\nonumber \=a_{f} &=& \epso \les  \=I - \le\eps_{a} \,\uz\uz
+\eps_{b}\,\ux\ux
\,  +\,\eps_{c}\,\uy\uy\ri\ris \\
&&\. \lec \=I + i \omega \epso \,\=D_{f} \. \les  \=I - \le\eps_{a}
\,\uz\uz +\eps_{b}\,\ux\ux \, +\,\eps_{c}\,\uy\uy\ri\ris
\ric^{-1}\,.
\end{eqnarray}
Herein, the depolarization
 dyadics
\begin{eqnarray}
\nonumber \=D_{s} &=&   \frac{1}{ i \omega \,\epso}
\,\frac{2}{\pi}\, \int_{\phi=0}^{\pi/2} \, d\phi
\int_{\theta=0}^{\pi/2}\, d\theta
\\
&&  \sin \theta \, \, \frac{ \frac{ \cos^{2}
\theta}{\gamma_{\tau}^{2}}
  \,\, \ux \ux+\sin^{2}
\theta\left( \cos^{2} \phi \,\, \uz\uz +  \frac{ \sin^{2}
\phi}{\gamma_{b}^{2}}  \,\, \uy\uy \right) }{\epsilon_{b}\,\,
\frac{\cos^{2} \theta}{\gamma_{\tau}^{2}} +
   \sin^{2} \theta \left(\epsilon_{a} \cos^{2} \phi +
\epsilon_{c} \, \frac{\sin^{2} \phi}{\gamma_{b}^{2}} \right)} \,,
\end{eqnarray}
and
\begin{eqnarray}
\nonumber \=D_{f} &=&   \frac{1}{ i \omega \,\epso}
\,\frac{2}{\pi}\, \int_{\phi=0}^{\pi/2} \, d\phi
\int_{\theta=0}^{\pi/2}\, d\theta
\\
&&  \sin \theta \, \, \frac{ { \cos^{2} \theta}{}
  \,\, \ux \ux+\sin^{2}
\theta\left( \cos^{2} \phi \,\, \uz\uz +  { \sin^{2} \phi}{}  \,\,
\uy\uy \right) }{\epsilon_{b}\,\, {\cos^{2} \theta}{} +
   \sin^{2} \theta \left(\epsilon_{a} \cos^{2} \phi +
\epsilon_{c} \, {\sin^{2} \phi}{} \right)}
\end{eqnarray}
are straightforwardly evaluated by numerical means.
 According to the Bruggeman homogenization formalism,
the three parameters $\lec n_s,f,\gamma_b\ric$  satisfy the three
nonlinear equations
\begin{equation}
b_\ell ( n_s, f, \gamma_b ) = 0, \qquad (\ell = x,y,z), \l{bj}
\end{equation}
where $b_\ell$ are the three nonzero components of the diagonal
dyadic $\=b$ ; i.e.,
\begin{equation}
\=b = b_x \, \ux\,\ux + b_y \, \uy\,\uy + b_z \, \uz\,\uz.
\end{equation}

Usually, the process of homogenization is applied in a forward
sense, to provide a nanoscopic-to-continuum model. Thereby,   the
relative permittivity parameters $\lec \eps_a,\eps_b,\eps_c\ric$ may
be estimated from a knowledge of the nanoscale parameters $\lec
n_s,f,\gamma_b\ric$. However,  the nanoscale parameters of CTFs are
generally unknown whereas $\lec \eps_a,\eps_b,\eps_c\ric$ may be
measured. In order to estimate $\lec n_s,f,\gamma_b\ric$
 from a knowledge of $\lec \eps_a,\eps_b,\eps_c\ric$, the inverse
homogenization process is needed. Formal expressions of the inverse
Bruggeman  formalism are available \c{WSW_MOTL}, but in certain
cases these formal expressions may be ill-defined \c{Cherkaev}.
In practice, it is more convenient to implement a direct numerical
method to compute $\lec n_s,f,\gamma_b\ric$, as described in the
next section.

\section{Numerical implementation}

 Solutions to Eqs.~\r{bj} may be computed  using a modified
 Newton--Raphson technique \c{Stark,Kampia}. In the recursive scheme
 implemented here, the estimated solutions at step $k+1$, namely
 $\lec n_s^{(k+1)},\,f^{(k+1)},\,\gamma_b^{(k+1)} \ric$, are derived
 from those at step $k$, namely $\lec n_s^{(k)},\,f^{(k)},\,\gamma_b^{(k)}
 \ric$, via
\begin{equation}
\left.
\begin{array}{l}
\displaystyle{n_s^{(k+1)} = n_s^{(k)} - \frac{b_x ( n_s^{(k)},
f^{(k)}, \gamma_b^{(k)}) }{\frac{\partial}{\partial n_s} b_x (
n_s^{(k)},
f^{(k)}, \gamma_b^{(k)})}} \vspace{6pt} \\
\displaystyle{f^{(k+1)} = f^{(k)} - \frac{b_y ( n_s^{(k+1)},
f^{(k)}, \gamma_b^{(k)}) }{\frac{\partial}{\partial f} b_y (
n_s^{(k+1)},
f^{(k)}, \gamma_b^{(k)})}} \vspace{6pt} \\
\displaystyle{\gamma_b^{(k+1)} = \gamma_b^{(k)} - \frac{b_z (
n_s^{(k+1)}, f^{(k+1)}, \gamma_b^{(k)}) }{\frac{\partial}{\partial
\gamma_b} b_z ( n_s^{(k+1)}, f^{(k+1)}, \gamma_b^{(k)})}}
\end{array} \l{NR1}
\right\}.
\end{equation}

In order for the  scheme \r{NR1} to converge, it is crucial that the
initial estimate
 $\lec n_s^{(0)},\,f^{(0)},\,\gamma_b^{(0)} \ric$ be sufficiently
 close to the true solution. A suitable initial estimate may be found
 by exploiting the forward Bruggeman formalism, as follows.

Let  $\tilde{\eps}_{a,b,c} $ denote  estimates of the CTF
permittivity parameters $\eps_{a,b,c}$, computed using the forward
Bruggeman formalism for physically reasonable ranges of the
parameters $n_s$, $f$ and $\gamma_b$,
 namely $ n_s \in
\le n_s^L, n_s^U \ri$, $f \in \le f^L, f^U \ri$ and  $ \gamma_b \in
\le \gamma_b^L, \gamma_b^U \ri$. Then:

\begin{itemize}
\item[(i)] Fix $ n_s = \le n_s^L + n_s^U \ri /2$ and  $ \gamma_b = \le
\gamma_b^L + \gamma_b^U \ri/2$. For all values of $f \in \le f^L,
f^U \ri$,  identify the value $f^*$  for which the quantity
 \begin{equation} \Delta = \sqrt{\le
\eps_a - \tilde{\eps}_a \ri^2 + \le \eps_b - \tilde{\eps}_b \ri^2 +
\le \eps_c - \tilde{\eps}_c \ri^2} \end{equation} is minimized.
\item[(ii)] Fix $f = f^*$ and  $ \gamma_b = \le \gamma_b^L + \gamma_b^U
\ri/2$. For all values of $n_s \in \le n_s^L, n_s^U \ri$, identify
the value $n_s^*$  for which $\Delta$ is minimized. \item[(iii)] Fix
$f = f^*$ and  $ n_s = n_s^*$. For all values of $\gamma_b \in \le
\gamma_b^L, \gamma_b^U \ri$, identify the value $\gamma_b^*$ for
which  $\Delta$ is minimized.
\end{itemize}
The steps (i)--(iii) are repeated, using $n_s^*$ and $\gamma_b^*$ as
the fixed values of $n_s$ and $\gamma_b$ in step (i), and
$\gamma_b^*$ as the fixed value of  $\gamma_b$ in step (ii), until
$\Delta$ becomes sufficiently small. In our numerical experiments,
we found that when $\Delta < 0.01$, the values of $n_s^*$, $f^*$ and
$\gamma_b^*$ provide suitable initial estimates for the modified
Newton--Raphson scheme \r{NR1}.

\section{Numerical results}
We considered CTFs made from three different materials: the oxides
of tantalum, titanium and zirconium. Experimental studies \c{HWH_AO}
have revealed that the permittivity parameters for these CTFs may be
expressed as
\begin{equation}
\label{Hodg1} \left. \begin{array}{ll}
\eps_{a}  =  \le n_{a0} + n_{a1} \,v + n_{a2} \,v^2 \ri^2\\[5pt]
\eps_{b} =  \le n_{b0} + n_{b1} \,v + n_{b2} \,v^2 \ri^2\\[5pt]
\eps_{c}  =  \le n_{c0} + n_{c1} \,v + n_{c2} \,v^2 \ri^2\\[5pt]
v =2\chi_v/\pi
\end{array}\ric ,
\end{equation}
wherein the vapor incidence angle $\chi_v \in(0,\pi/2]$ is related
to the column inclination angle by the coefficient $\bar{m}$, per
\begin{equation}
\tan \chi = \bar{m} \, \tan \chi_v .
\end{equation}

Table 1 contains values of the  ten coefficients $n_{a0}$
to $\bar{m}$    of CTFs of the three different materials.   Although the bulk refractive indexes
of all three oxides are quite close to each other, the coefficients
$n_{a0}$
to $m$ of the three types of CTFs are quite different, as indeed are also their
constitutive parameters $\epsilon_{a,b,c}$ \cite{Chiadini}. These differences arise, in significant measure,
due to the dependence of the
growth dynamics of a CTF on the evaporated bulk material \cite{MVS,Mess08}.

\begin{table}[h]
\caption{ Coefficients appearing in Eqs. \r{Hodg1}, obtained from
the experimental findings of Hodgkinson {\em et al.\/} \cite{HWH_AO}
on CTFs, when the free-space wavelength is $633$~nm.}
\begin{center}
\begin{tabular}{|c||c|c|c||c|c|c|}
\hline
material & $n_{a0}$ & $n_{a1}$ & $n_{a2}$
              & $n_{b0}$ & $n_{b1}$ & $n_{b2}$
               \\
\hline\hline
tantalum & 1.1961 & 1.5439 & $-$0.7719
               & 1.4600 & 1.0400 & $-$0.5200
               \\
oxide & & & & & &   \\
\hline
titanium & 1.0443 & 2.7394& $-$1.3697
               & 1.6765 & 1.5649 & $-$0.7825
               \\
oxide & & & & & &   \\
\hline
zirconium & 1.2394 & 1.2912& $-$0.6456
               & 1.4676 & 0.9428 & $-$0.4714
               \\
oxide & & & & & &    \\
\hline

\end{tabular}

\vskip 5pt

\hskip-13pt
\begin{tabular}{|c||c|c|c||c|}
\hline
material   & $n_{c0}$ & $n_{c1}$ & $n_{c2}$
              & $\bar{m}$\\
\hline\hline
tantalum & 1.3532 & 1.2296 & $-$0.6148
        & 3.1056\\
oxide   & & &  & \\
\hline
titanium & 1.3586 & 2.1109 & $-$1.0554
        & 2.8818\\
oxide   & & &  & \\
\hline
zirconium & 1.3861 & 0.9979 & $-$0.4990
        & 3.5587\\
oxide   & & &  &  \\
\hline

\end{tabular}
\end{center}
\end{table}

\vskip 20pt

\begin{figure}[!ht]
\centering
\includegraphics[width=3.5in]{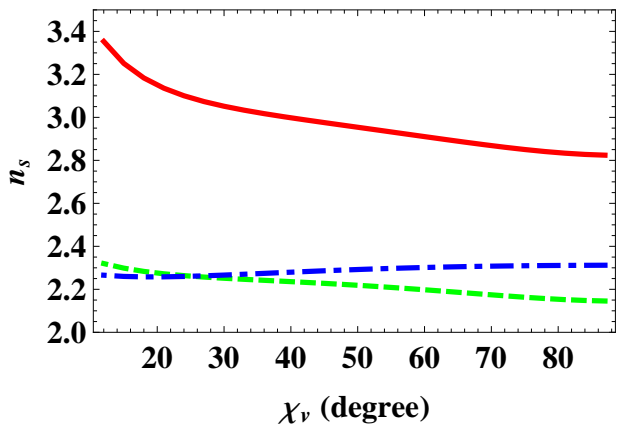}
 \caption{ \l{Fig1}
 The quantity $n_s$ plotted against $\chi_v$ (in degree)
 for CTFs made from evaporating  titanium oxide (red, solid curve), tantalum oxide (green, dashed curve) and zirconium oxide (blue, broken dashed curve), as computed using the inverse Bruggeman formalism.}
\end{figure}

\begin{figure}[!ht]
\centering
\includegraphics[width=3.5in]{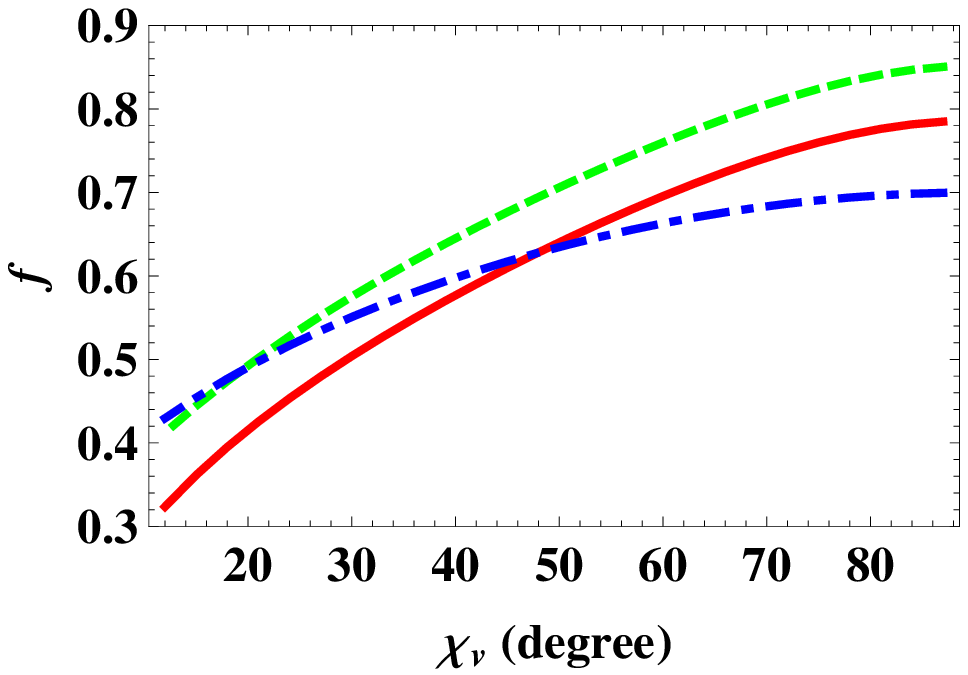}
 \caption{ \l{Fig2}
As Fig.~\ref{Fig1} except that the quantity plotted against $\chi_v$
 is $f$.}
\end{figure}

\begin{figure}[!ht]
\centering
\includegraphics[width=3.5in]{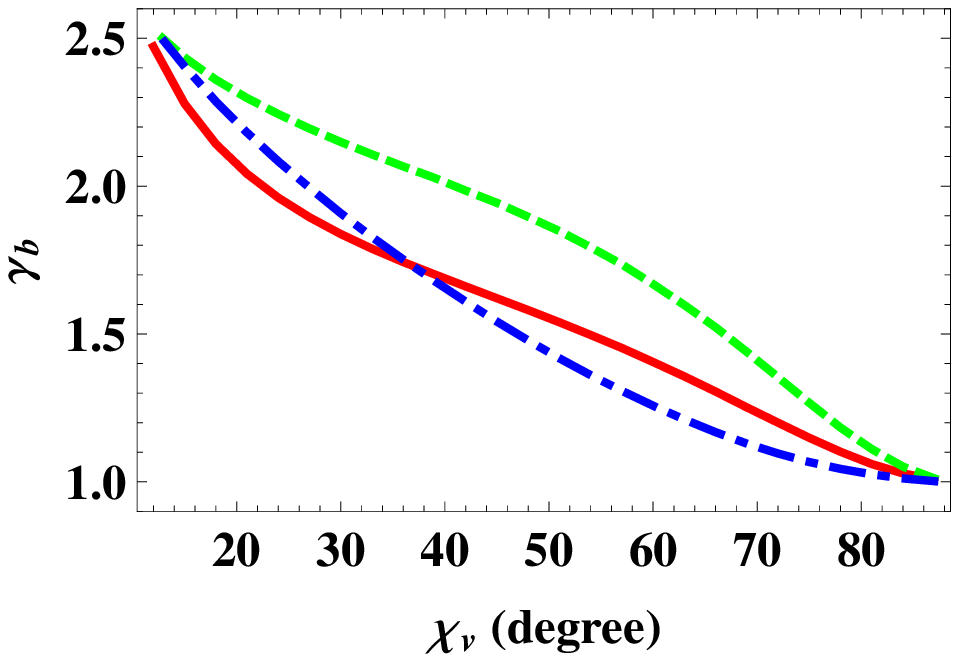}
 \caption{ \l{Fig3}
As Fig.~\ref{Fig1} except that the quantity plotted against $\chi_v$
is $\gamma_b$.}
\end{figure}

The nanoscale parameters $\lec n_s,f,\gamma_b\ric$ were estimated for the three
CTFs using the modified Newton--Raphson technique \r{NR1}, with
initial guesses $\lec n_s^{(0)},\,f^{(0)},\,\gamma_b^{(0)} \ric$
deduced by scanning the solution space of
$\lec \tilde{\eps}_a,\tilde{\eps}_b,\tilde{\eps}_c\ric$
with $n_s^L = 1$, $n_s^U = 4$, $f^L = 0.2$, $f^U = 0.9$, $\gamma_b^L
= 0.5$ and $\gamma_b^U = 3$.

The computed nanoscale
parameters $n_s$, $f$, and $\gamma_b$, respectively, are plotted in Figs.~\ref{Fig1}--\ref{Fig3},
against $\chi_v \in (12^\circ, 90^\circ)$ for the CTFs fabricated by evaporating any one of the  three
bulk materials. The plots in
Fig.~\ref{Fig1}  show that
$n_s$  for CTFs made by evaporating any of the three bulk materials
to be largely insensitive to $\chi_v$.
In contrast,
the volume fractions $f$ displayed in Fig.~\ref{Fig2} for all three
materials increase rapidly as $\chi_v$ increases, in general accord with
the observation that mass density of a CTF varies as $(1+\sin\chi_v)^{-1}\sin\chi_v$ \cite{TSB}.
The shape
parameters $\gamma_b$ displayed in Fig.~\ref{Fig3} for CTFs of all three
evaporated bulk materials decrease rapidly towards unity as $\chi_v$ increases.
This is in accord with the observation that CTFs deposited with $\chi_v=90^\circ$
are macroscopically uniaxial rather than biaxial \cite{HW}.

\section{Concluding remarks}

In order to exploit the considerable potential that CTFs possess for widespread applications such as
optical sensors of analytes, it is
vital that they be
reliably characterized at the nanoscale. Our theoretical and
numerical study has demonstrated that the
 inverse Bruggeman homogenization formalism
 provides a practicable  means for  this characterization, in terms
of three nanoscale parameters. Thus, a key step towards the intelligent design and development of
CTF-based (and other STF-based \cite{AZL}) optical sensors has been taken.

\vspace{10mm}

\noindent {\bf Acknowledgments:} TGM is supported by a  Royal
Academy of Engineering/Leverhulme Trust Senior Research Fellowship.
AL thanks the Binder Endowment at Penn State for partial financial
support of his research activities.

\end{document}